\documentclass[fleqn,usenatbib,useAMS]{mnras}
 \usepackage{graphicx}
 \usepackage{amsmath}
 \usepackage{amssymb}
 \usepackage[T1]{fontenc}
 \usepackage{ae,aecompl}
 \usepackage{mathptmx}
 \usepackage{txfonts}
 \usepackage{lscape}

 \title[A recurring co-orbital to Uranus]
       {Asteroid 2015~DB$_{216}$: a recurring co-orbital companion to Uranus}

 \author[C. de la Fuente Marcos and R. de la Fuente Marcos]
        {C.~de~la~Fuente~Marcos\thanks{E-mail: carlosdlfmarcos@gmail.com}
         and
         R. de la Fuente Marcos \\
         Apartado de Correos 3413, E-28080 Madrid, Spain}
 \date{Accepted 2015 July 27.
       Received 2015 July 7;
       in original form 2015 June 3}
 \pubyear{2015}
 
\begin{document}
  \label{firstpage}
  \pagerange{\pageref{firstpage}--\pageref{lastpage}}
  \maketitle

  \begin{abstract}
     Minor bodies trapped in 1:1 co-orbital resonances with a host planet could 
     be relevant to explain the origin of captured satellites. Among the giant 
     planets, Uranus has one of the smallest known populations of co-orbitals, 
     three objects, and all of them are short-lived. Asteroid 2015~DB$_{216}$  
     has an orbital period that matches well that of Uranus, and here we 
     investigate its dynamical state. Direct $N$-body calculations are used to 
     assess the current status of this object, reconstruct its immediate 
     dynamical past, and explore its future orbital evolution. A covariance 
     matrix-based Monte Carlo scheme is presented and applied to study its 
     short-term stability. We find that 2015~DB$_{216}$ is trapped in a 
     temporary co-orbital resonance with Uranus, the fourth known minor body to 
     do so. A detailed analysis of its dynamical evolution shows that it is an 
     unstable but recurring co-orbital companion to Uranus. It currently follows 
     an asymmetric horseshoe trajectory that will last for at least 10 kyr, but 
     it may remain inside Uranus' co-orbital zone for millions of years. As in 
     the case of other transient Uranian co-orbitals, complex multibody 
     ephemeral mean motion resonances trigger the switching between the various 
     resonant co-orbital states. The new Uranian co-orbital exhibits a secular 
     behaviour markedly different from that of the other known Uranian 
     co-orbitals because of its higher inclination, nearly 38\degr. Given its 
     rather unusual discovery circumstances, the presence of 2015~DB$_{216}$ 
     hints at the existence of a relatively large population of objects moving 
     in similar orbits. 
  \end{abstract}

  \begin{keywords}
     methods: numerical -- methods: statistical -- celestial mechanics --
     minor planets, asteroids: individual: 2015~DB$_{216}$ --
     planets and satellites: individual: Uranus.
  \end{keywords}

  \section{Introduction}
     For over a century, co-orbitals ---or minor bodies trapped in a 1:1 mean motion resonance with a host planet--- have been regarded as 
     mere interesting dynamical curiosities (Jackson 1913; Henon 1969; Namouni 1999). This view is now changing considerably; in fact, the 
     heliocentric 1:1 co-orbital resonance could be an efficient mechanism for capture of satellites by a planet and therefore explain the 
     origin of some irregular moons (Kortenkamp 2005). This theoretical possibility ---that of being a feasible dynamical pathway to capture 
     satellites, at least temporarily--- was dramatically vindicated when a co-orbital of the Earth, 2006~RH$_{120}$, remained as natural 
     satellite of our planet for about a year starting in 2006 June (Kwiatkowski et al. 2009; Granvik, Vaubaillon \& Jedicke 2012).
     \hfil\par 
     Uranus has one of the smallest known populations of co-orbitals and all of them are relatively short-lived (de la Fuente Marcos \& de 
     la Fuente Marcos 2014). So far, Uranus had only three known co-orbitals: 83982 Crantor (2002 GO$_{9}$) (Gallardo 2006; de la Fuente 
     Marcos \& de la Fuente Marcos 2013), 2010 EU$_{65}$ (de la Fuente Marcos \& de la Fuente Marcos 2013), and 2011~QF$_{99}$ (Alexandersen 
     et al. 2013; de la Fuente Marcos \& de la Fuente Marcos 2014). Due to its present short data-arc (85 d), 2010 EU$_{65}$ is better 
     described as a candidate. Asteroids Crantor and 2010 EU$_{65}$ follow horseshoe orbits, and 2011~QF$_{99}$ is an L$_4$ Trojan. 
     Consistently, Uranus has also a small population of irregular satellites, significantly smaller than that of Jupiter or Saturn (Grav et 
     al. 2003; Sheppard, Jewitt \& Kleyna 2005). Most Uranian irregular satellites are retrograde and their large spread in semimajor axis
     suggest that they formed independently (Nesvorn\'y et al. 2003); orbits with inclinations in the range (80\degr, 100\degr) are unstable 
     due to the Kozai resonance (Kozai 1962). For the particular case of Uranus' Trojans, Dvorak, Bazs\'o \& Zhou (2010) have also found that
     the stability depends on the orbital inclination and only the inclination intervals (0\degr, 7\degr), (9\degr, 13\degr), (31\degr, 
     36\degr), and (38\degr, 50\degr) seem to be stable. Asteroid 2011~QF$_{99}$ appears to inhabit one of these stable islands at an 
     inclination of nearly 11\degr (de la Fuente Marcos \& de la Fuente Marcos 2014). The stability of Uranian Trojans had been previously 
     studied by Marzari, Tricarico \& Scholl (2003), and by Nesvorn\'y \& Dones (2002) and Holman \& Wisdom (1993) before them.
     \hfil\par
     Here, we present a recently discovered object, 2015~DB$_{216}$, that is also trapped in a 1:1 mean motion resonance with Uranus. This
     minor body exhibits some dynamical features that separate it from the previously known Uranian co-orbitals. This paper is organized as 
     follows. In Section 2, we briefly discuss both the data and the numerical model used in our calculations. The topic of generating 
     control orbits compatible with the available observations and its implications is considered in Section 3. The current status of 
     2015~DB$_{216}$ is studied in Section 4, where its dynamical past and future orbital evolution are also investigated. Section 5 
     discusses our results and their possible significance. The stability of the co-orbital realm located in the neighbourhood of 
     2015~DB$_{216}$ is tentatively explored in Section 6. A summary of our conclusions is given in Section 7.

  \section{Data and methodology}
     Asteroid 2015~DB$_{216}$ was discovered on 2015 February 27 at Mt. Lemmon Survey. With a value of the semimajor axis $a$ = 19.20 au,
     this Centaur moves in an eccentric, $e$ = 0.32, and highly inclined path, $i$ = 37\fdg72. With such an orbit, close encounters are 
     only possible with Uranus as its perihelion is well beyond Saturn's aphelion and its aphelion far from Neptune's perihelion. It is a 
     relatively large object with $H$ = 8.3 mag which translates into a diameter in the range 46--145 km for an assumed albedo of 
     0.40--0.04. Its period of revolution around the Sun, approximately 84.16 yr at present, is very close to that of Uranus which is 
     suggestive of an object that moves co-orbital with the giant planet. Its current orbit is statistically robust because six precovery
     images acquired by the Sloan Digital Sky Survey (SDSS) at Apache Point late in 2003 have been found. The heliocentric Keplerian 
     osculating orbital elements and uncertainties in Table \ref{elements} are based on 28 observations for a data-arc span of 4\,200 d and 
     they have been obtained from the Jet Propulsion Laboratory (JPL) Small-Body Database.\footnote{http://ssd.jpl.nasa.gov/sbdb.cgi} 
%
%
         \begin{table}
          \fontsize{8}{11pt}\selectfont
          \tabcolsep 0.30truecm
          \caption{Heliocentric Keplerian orbital elements of 2015~DB$_{216}$ used in this research. The orbit is based on 28 observations 
                   spanning a data-arc of 4\,200 days or 11.50 yr, from 2003 October 21 to 2015 April 23. Values include the 1$\sigma$ 
                   uncertainty. The orbit is computed at epoch JD 2457000.5 that corresponds to 0:00 UT on 2014 December 9 (J2000.0 ecliptic 
                   and equinox) and it is $t = 0$ in the figures. Source: JPL Small-Body Database. 
                  }
          \begin{tabular}{ccc}
           \hline
            Semimajor axis, $a$ (au)                         & = & 19.204$\pm$0.005 \\
            Eccentricity, $e$                                 & = & 0.32395$\pm$0.00013 \\
            Inclination, $i$ (\degr)                          & = & 37.7173$\pm$0.0003 \\
            Longitude of the ascending node, $\Omega$ (\degr) & = & 6.2679$\pm$0.0003 \\
            Argument of perihelion, $\omega$ (\degr)          & = & 237.75$\pm$0.03 \\
            Mean anomaly, $M$ (\degr)                         & = & 302.52$\pm$0.04 \\
            Perihelion, $q$ (au)                              & = & 12.9832$\pm$0.0013 \\
            Aphelion, $Q$ (au)                                & = & 25.426$\pm$0.007 \\
            Absolute magnitude, $H$ (mag)                     & = & 8.3$\pm$0.4 \\
           \hline
          \end{tabular}
          \label{elements}
         \end{table}
%
%
     \hfil\par
     In order to assess the dynamical status of 2015~DB$_{216}$, we use the Hermite integration scheme described by Makino (1991) and 
     implemented by Aarseth (2003). The standard version of this direct $N$-body code is publicly available from the IoA web 
     site.\footnote{http://www.ast.cam.ac.uk/$\sim$sverre/web/pages/nbody.htm} Our physical model includes the perturbations by the eight 
     major planets, the Moon, the barycentre of the Pluto-Charon system, and the three largest asteroids; additional details can be found in
     de la Fuente Marcos \& de la Fuente Marcos (2012). To compute accurate initial positions and velocities we used the heliocentric 
     ecliptic Keplerian elements provided by the JPL On-line Solar System Data Service\footnote{http://ssd.jpl.nasa.gov/?planet\_pos} 
     (Giorgini et al. 1996) and initial positions and velocities based on the DE405 planetary orbital ephemerides (Standish 1998) referred 
     to the barycentre of the Solar system. Besides the orbital calculations completed using the nominal elements in Table \ref{elements}, 
     we have performed 50 control simulations with sets of orbital elements obtained from the nominal ones as described in the following 
     section, all of them for 0.5 Myr forward and backwards in time. Two more sets of 100 control orbits each have been integrated for just 
     5 kyr into the past and the future to better characterize its short-term stability.

  \section{Generating control orbits with Monte Carlo and the covariance matrix}
     For a given minor body, the orbital elements are a coordinate in six-dimensional space (assuming as we do that non-gravitational forces 
     can be neglected), which represents the location where samples of control orbits are most likely to be generated. This is analogous to 
     the peak of the Gaussian curve for a typical one-dimensional or univariate normal distribution. The multivariate normal distribution is
     a generalization of the one-dimensional normal distribution to higher dimensions. Instead of being specified by its mean value and 
     variance, such a distribution is characterized by its mean (a vector with the mean of the multidimensional distribution) and 
     covariance matrix, which defines an hyperellipsoid in multidimensional space. The values of the elements of the covariance matrix 
     indicate the level to which two given variables vary together. For a particular object, both mean and covariance matrix are computed 
     from the observations. 
     \hfil\par
     When studying the stability of the orbital solution of a certain minor planet, we can compute the orbital elements of the control 
     orbits varying them randomly, within the ranges defined by their mean values and standard deviations. For example, a new value of 
     the orbital eccentricity can be found using the expression $e_{\rm c} = e + \sigma_{\rm e}\,r_{\rm i}$, where $e_{\rm c}$ is the 
     eccentricity of the control orbit, $e$ is the mean value of the eccentricity (nominal orbit), $\sigma_{\rm e}$ is the standard 
     deviation of $e$ (nominal orbit), and $r_{\rm i}$ is a (pseudo) random number with normal distribution in the range $-$1 to 1. In 
     statistical terms, the univariate Gaussian distribution results from adding a standard Gaussian variate with mean zero and variance 
     one multiplied by the standard deviation, to the mean value. This is equivalent to considering a number of different virtual minor 
     planets moving in similar orbits, not a sample of control orbits incarnated from a set of observations obtained for a single minor 
     planet. If the control orbits are meant to be compatible with actual observations, we have to consider how the elements affect each 
     other using the covariance matrix or e.g. following the procedure described in Sitarski (1998, 1999, 2006).
     \hfil\par
\begin{figure}
        \centering
         \includegraphics[width=0.49\linewidth]{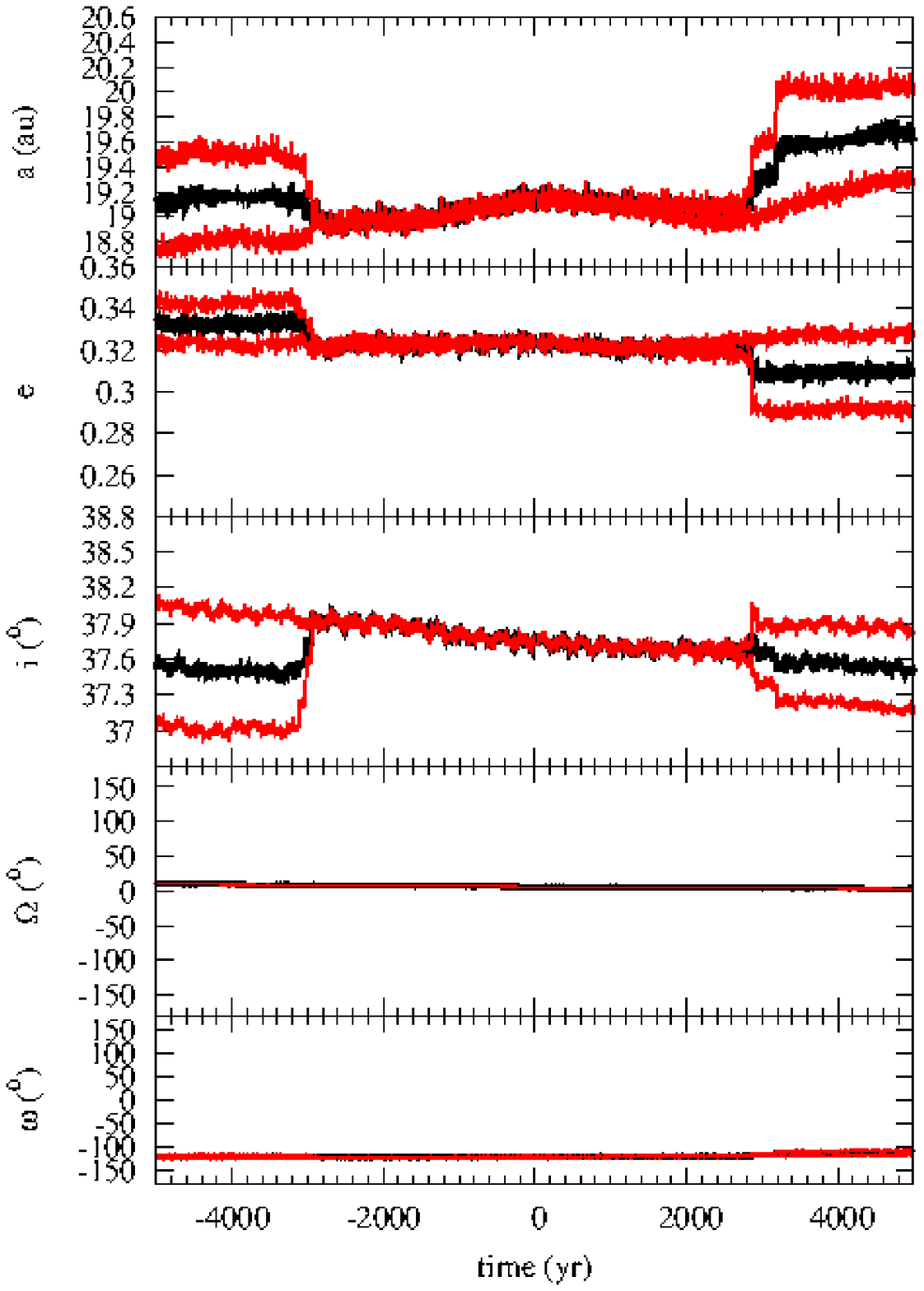}
         \includegraphics[width=0.49\linewidth]{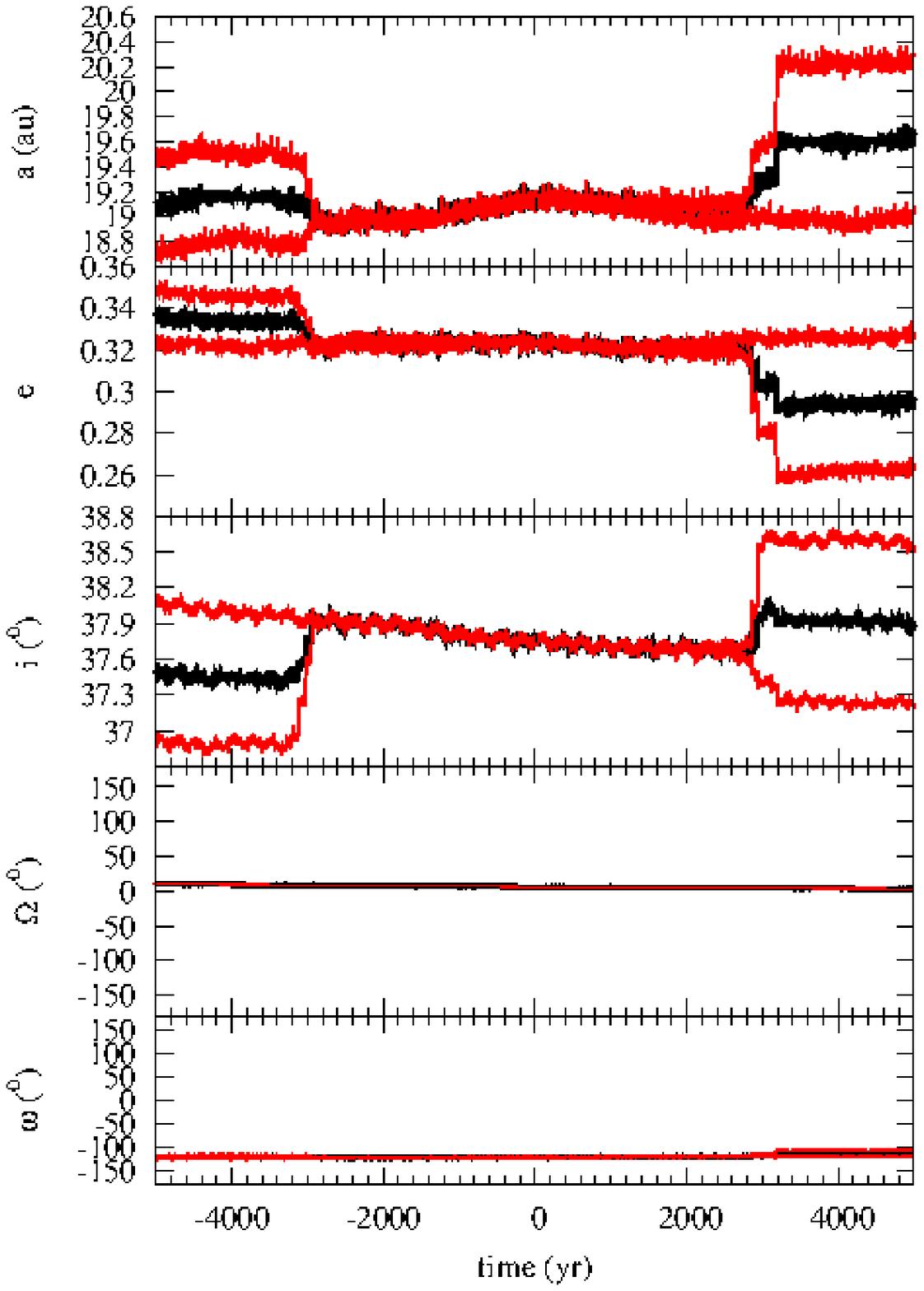}
         \caption{Time evolution of the orbital elements $a$, $e$, $i$, $\Omega$, and $\omega$ of 2015~DB$_{216}$. The thick black curve 
                  shows the average evolution of 100 control orbits, the thin red curves display the ranges in the values of the parameters 
                  at the given time. Control orbits depicted in the left-hand panels have been computed as described in the text, using the 
                  covariance matrix (Eqs. \ref{good}). Those displayed in the right-hand panels have been computed without taking into 
                  account the covariance matrix (Eqs. \ref{bad}).
                 }
         \label{disper}
      \end{figure}

     The methodology used in this paper is an implementation of the classical Monte Carlo using the Covariance Matrix (MCCM, Bordovitsyna, 
     Avdyushev \& Chernitsov 2001; Avdyushev \& Banschikova 2007) approach, i.e. a Monte Carlo process creates control orbits with initial 
     parameters from the nominal orbit adding random noise on each initial orbital element making use of the covariance matrix. The MCCM
     approach considers that the estimated parameters are Gaussian random variables with mean values those of the nominal orbit and 
     covariance matrix obtained via the least-squares method applied to the available observations. Assuming a covariance matrix as computed 
     by the JPL Solar System Dynamics Group, Horizons On-Line Ephemeris System, the vector including the mean values of the orbital 
     parameters at a given epoch $t_{0}$ is of the form $\textit{\textbf{v}} = (e, q, \tau, \Omega, \omega, i)$; the perihelion is given by 
     the expression $q = a\,(1-e)$. If \textbf{\textsf{C}} is the covariance matrix at the same epoch associated with the nominal orbital 
     solution that is symmetric and positive-semidefinite, then \textbf{\textsf{C}} = \textbf{\textsf{A}} \textbf{\textsf{A}}$^{\textbf{\textsf{T}}}$, 
     where \textbf{\textsf{A}} is a lower triangular matrix with real and positive diagonal elements, \textbf{\textsf{A}}$^{\textbf{\textsf{T}}}$ 
     is the transpose of \textbf{\textsf{A}}. In the particular case studied here, these matrices are $6\times6$. If the elements of 
     \textbf{\textsf{C}} are written as $c_{\rm ij}$ and those of \textbf{\textsf{A}} as $a_{\rm ij}$, where those are the entries in the 
     $i$-th row and $j$-th column, they are related by the following expressions: 
     \begin{equation}
        \begin{aligned}
           a_{11} & = \sqrt{c_{11}} \\
           a_{21} & = c_{12} / a_{11} \\
           a_{31} & = c_{13} / a_{11} \\
           a_{41} & = c_{14} / a_{11} \\
           a_{51} & = c_{15} / a_{11} \\
           a_{61} & = c_{16} / a_{11} \\
           a_{22} & = \sqrt{c_{22} - a_{21}^{2}} \\
           a_{32} & = (c_{23} - a_{21}\,a_{31}) / a_{22} \\
           a_{42} & = (c_{24} - a_{21}\,a_{41}) / a_{22} \\
           a_{52} & = (c_{25} - a_{21}\,a_{51}) / a_{22} \\
           a_{62} & = (c_{26} - a_{21}\,a_{61}) / a_{22} \\
           a_{33} & = \sqrt{c_{33} - a_{31}^{2} - a_{32}^{2}} \\
           a_{43} & = (c_{34} - a_{31}\,a_{41} - a_{32}\,a_{42}) / a_{33} \\
           a_{53} & = (c_{35} - a_{31}\,a_{51} - a_{32}\,a_{52}) / a_{33} \\
           a_{63} & = (c_{36} - a_{31}\,a_{61} - a_{32}\,a_{62}) / a_{33} \\
           a_{44} & = \sqrt{c_{44} - a_{41}^{2} - a_{42}^{2} - a_{43}^{2}} \\
           a_{54} & = (c_{45} - a_{41}\,a_{51} - a_{42}\,a_{52} - a_{43}\,a_{53}) / a_{44} \\
           a_{64} & = (c_{46} - a_{41}\,a_{61} - a_{42}\,a_{62} - a_{43}\,a_{63}) / a_{44} \\
           a_{55} & = \sqrt{c_{55} - a_{51}^{2} - a_{52}^{2} - a_{53}^{2} - a_{54}^{2}} \\
           a_{65} & = (c_{56} - a_{51}\,a_{61} - a_{52}\,a_{62} - a_{53}\,a_{63} - a_{54}\,a_{64}) / a_{55} \\
           a_{66} & = \sqrt{c_{66} - a_{61}^{2} - a_{62}^{2} - a_{63}^{2} - a_{64}^{2} - a_{65}^{2}} \,.
        \end{aligned}
     \end{equation}
     If \textit{\textbf{r}} is a vector made of univariate Gaussian random numbers (components $r_{\rm i}$ with $i=1,6$), the required 
     multivariate Gaussian random samples ---i.e. the sets of initial orbital elements of the control orbits--- are given by the 
     expressions (assuming the structure provided by the JPL Horizons On-Line Ephemeris System), $\textit{\textbf{v}}_{c} = 
     \textit{\textbf{v}} + \textbf{\textsf{A}}\,\textit{\textbf{r}}$:
     \begin{equation}
        \begin{aligned}
           e_{\rm c} & = e + a_{11}\,r_{1} \\
           q_{\rm c} & = q + a_{22}\,r_{2} + a_{21}\,r_{1} \\
           \tau_{\rm c} & = \tau + a_{33}\,r_{3} + a_{32}\,r_{2} + a_{31}\,r_{1} \\
           \Omega_{\rm c} & = \Omega + a_{44}\,r_{4} + a_{43}\,r_{3} + a_{42}\,r_{2} + a_{41}\,r_{1} \\
           \omega_{\rm c} & = \omega + a_{55}\,r_{5} + a_{54}\,r_{4} + a_{53}\,r_{3} + a_{52}\,r_{2} + a_{51}\,r_{1} \\
           i_{\rm c} & = i + a_{66}\,r_{6} + a_{65}\,r_{5} + a_{64}\,r_{4} + a_{63}\,r_{3} + a_{62}\,r_{2} + a_{61}\,r_{1} \,. 
           \label{good}
        \end{aligned}
     \end{equation}
     In contrast, the equivalent classical ---but statistically wrong--- expressions commonly used to generate control orbits are given 
     by:
     \begin{equation}
        \begin{aligned}
           e_{\rm c} & = e + \sigma_{e}\,r_{1} \\
           q_{\rm c} & = q + \sigma_{q}\,r_{2} \\
           \tau_{\rm c} & = \tau + \sigma_{\tau}\,r_{3} \\
           \Omega_{\rm c} & = \Omega +  \sigma_{\Omega}\,r_{4} \\
           \omega_{\rm c} & = \omega +  \sigma_{\omega}\,r_{5} \\
           i_{\rm c} & = i +  \sigma_{i}\,r_{6} \,.
           \label{bad}
        \end{aligned}
     \end{equation}
     A comparison between the results of the evolution of a sample of control orbits generated using Eqs. \ref{good} and \ref{bad} for
     the particular case of 2015~DB$_{216}$ appears in Fig. \ref{disper}, left-hand and right-hand panels, respectively. In our 
     calculations, the Box-Muller method (Press et al. 2007) was used to generate random numbers with a normal distribution. It is obvious 
     that, at least for this particular object, the difference is not very significant. However and for very precise orbits, the outcomes 
     from these two approaches could be very different (see e.g. fig. 5 in Sitarski 1998); creating control orbits by randomly varying the 
     nominal orbital elements in range of their mean errors (Eqs. \ref{bad}) is not recommended in that case.
%
%
%
%

  \section{Asteroid 2015~DB$_{216}$: dynamical evolution}
     In order to assess the dynamical status of 2015~DB$_{216}$, we focus on the study of the librational behaviour of the relative mean 
     longitude $\lambda_{\rm r} = \lambda - \lambda_{\rm U}$, where $\lambda$ and $\lambda_{\rm U}$ are the mean longitudes of the object 
     and Uranus, respectively; $\lambda$ = $M$ + $\Omega$ + $\omega$, where $M$ is the mean anomaly, $\Omega$ is the longitude of the 
     ascending node, and $\omega$ is the argument of perihelion. If $\lambda_{\rm r}$ oscillates around 0\degr, the object is considered a 
     quasi-satellite; Trojan bodies are characterized by $\lambda_{\rm r}$ librating around +60\degr (L$_4$ Trojan) or $-$60\degr (or 
     300\degr, L$_5$ Trojan); finally, an object whose $\lambda_{\rm r}$ oscillates with amplitude $>$~180\degr follows a horseshoe orbit 
     (see e.g. Murray \& Dermott 1999). Quasi-satellites are not true gravitationally bound satellites but appear to orbit the host planet 
     like a retrograde satellite. If $\lambda_{\rm r}$ can take any value (circulates), we speak of passing orbits.
%
%
      \begin{figure}
        \centering
         \includegraphics[width=\linewidth]{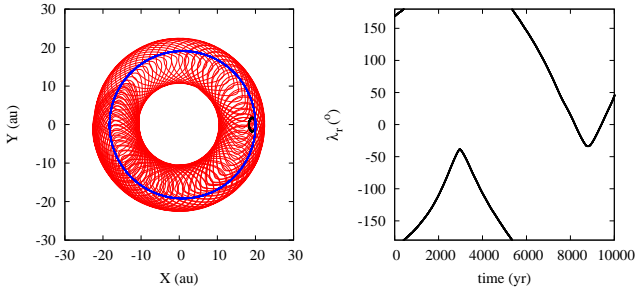}
         \caption{The motion of 2015~DB$_{216}$ during the time interval (0,~10) kyr projected on to the ecliptic plane in a coordinate 
                  system rotating with Uranus (red curve, left-hand panel). The orbit and the position of Uranus are also plotted (blue 
                  curve). In this frame of reference, and as a result of its non-negligible eccentricity, Uranus describes a small ellipse 
                  (black curve). The associated values of the resonant angle, $\lambda_{\rm r}$, are also displayed (right-hand panel). 
                 }
         \label{two}
      \end{figure}
%
%
     \hfil\par
\begin{figure*}
        \centering
         \includegraphics[width=\linewidth]{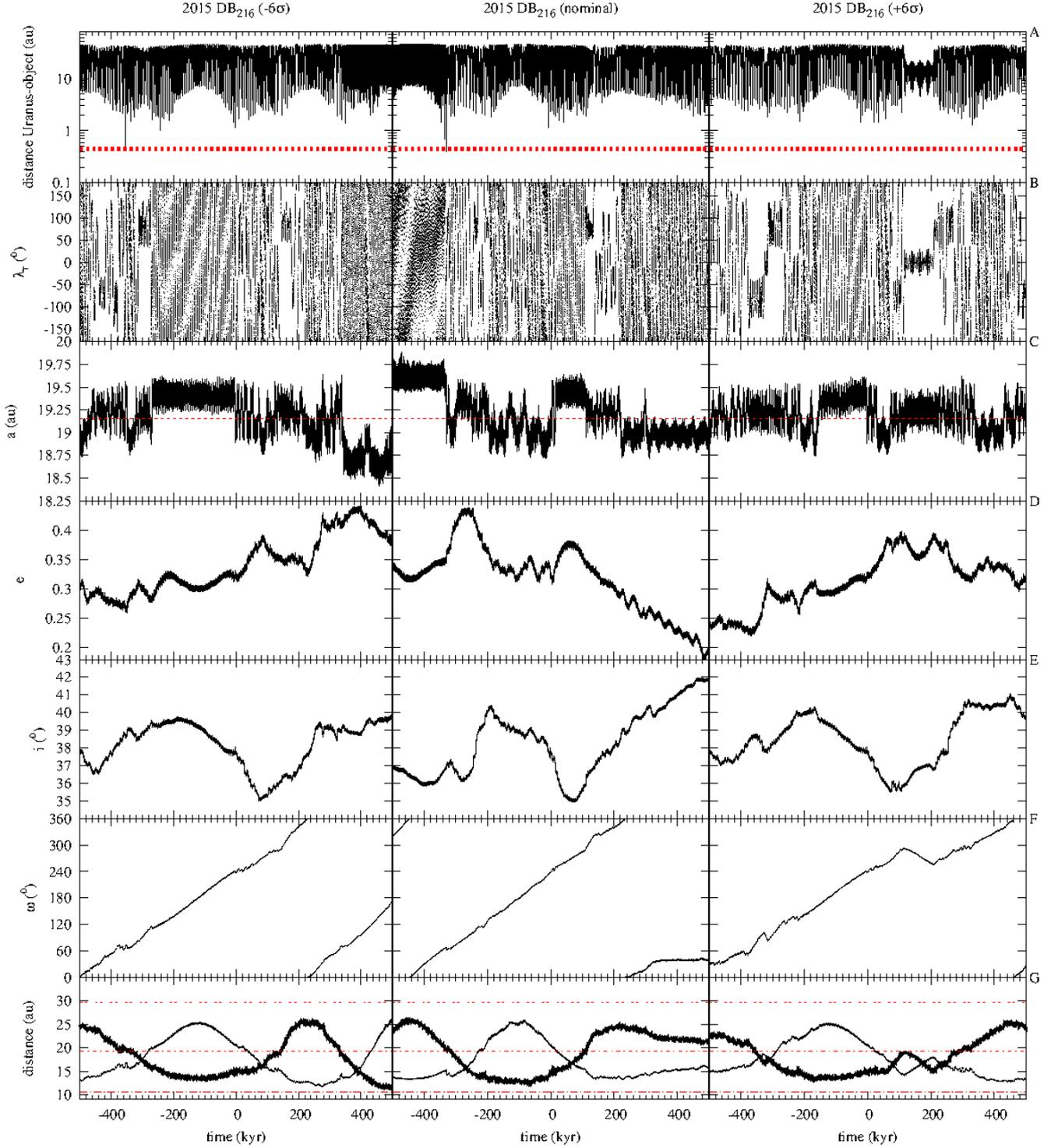}
         \caption{Comparative dynamical evolution of various parameters for the nominal orbit of 2015~DB$_{216}$ as presented in Table 
                  \ref{elements} (central panels) and two representative orbits that are most different from the nominal one (see the text 
                  for details). The distance from Uranus (A-panels); the value of the Hill sphere radius of Uranus, 0.4469 au, is displayed 
                  as a red line. The resonant angle, $\lambda_{\rm r}$ (B-panels). The orbital elements $a$ (C-panels, the value of the 
                  semimajor axis of Uranus appears as a red line), $e$ (D-panels), $i$ (E-panels) and $\omega$ (F-panels). The distances to 
                  the descending (thick line) and ascending nodes (dotted line) appear in the G-panels. Saturn and Neptune aphelion and 
                  perihelion distances are also shown as red lines.
                 }
         \label{controldb216}
      \end{figure*}

 
     Our $N$-body integrations show that 2015~DB$_{216}$ is currently a co-orbital companion to Uranus and moves in an asymmetric horseshoe 
     orbit with a period of about 11 kyr (see Fig. \ref{two}, right-hand panel); in this case, asymmetric means that the resonant angle, 
     $\lambda_{\rm r}$, goes beyond 0\degr reaching an offset of libration around $-$40\degr at nearly 9 kyr. The left-hand panel in Fig. 
     \ref{two} depicts the trajectory of 2015~DB$_{216}$ viewed in a frame of reference corotating with Uranus. Figure \ref{controldb216} 
     displays the dynamical evolution of various parameters for three representative orbits: the nominal one (central panels) and those of 
     two additional orbits which are most different from the previous one, and have been obtained adding (+) or subtracting ($-$) 6-times 
     the uncertainty from the orbital parameters (the six elements) in Table \ref{elements}. All the control orbits show consistent 
     behaviour within a few thousand years of $t = 0$ (see Fig. \ref{disper}). Its $e$-folding time, or characteristic time-scale on which 
     two arbitrarily close orbits diverge exponentially, is a few thousand years both in the past and the future. The evolution of the 
     control orbits exhibits very similar behaviour of all the orbital elements within the time frame ($-$3, 3) kyr (see Fig. \ref{disper}).
%
%
%
%
     \hfil\par
     Asteroid 2015~DB$_{216}$ currently occupies (see Fig. \ref{controldb216}, E-panels) a band of instability between the two stable 
     islands in inclination, (31\degr, 36\degr) and (38\degr, 50\degr), described in Dvorak et al. (2010) for Uranian Trojans. However, the 
     figure shows that the inclination of this asteroid is high enough to avoid close encounters with Uranus when the relative mean 
     longitude approaches zero i.e. close encounters with Uranus (or any other body) are not responsible for the activation and deactivation 
     of the co-orbital behaviour of this object. Very few close encounters with Uranus have been observed during the simulated time 
     (examples appear in the A-left-hand and central panels of Fig. \ref{controldb216}). However, multiple and repetitive short co-orbital 
     episodes of the Trojan, quasi-satellite and horseshoe type are observed in Fig. \ref{controldb216}. Recurrent co-orbital episodes in 
     which the relative mean longitude librates for several cycles and then circulates for a few more cycles before restarting libration 
     once again, are the signpost of a type of dynamical behaviour known as resonance angle nodding, see e.g. Ketchum, Adams \& Bloch 
     (2013); nodding often occurs when a small body is in an external (near) mean motion resonance with a larger planet. In our case, the 
     situation is more complicated because we have multiple distant perturbers. 
     \hfil\par
\begin{figure}
        \centering
         \includegraphics[width=\linewidth]{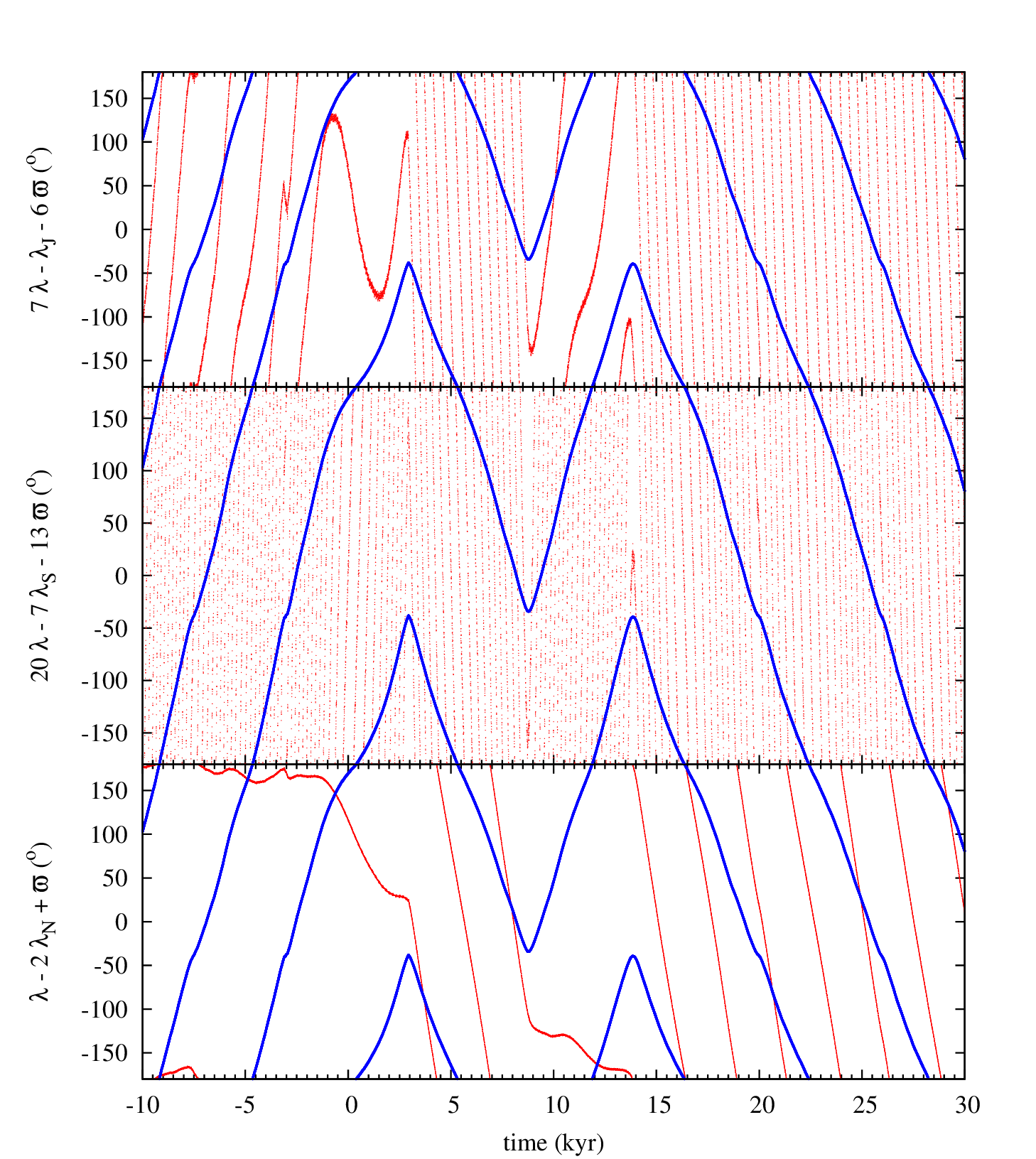}
         \caption{Resonant arguments $\sigma_{\rm J} = 7 \lambda - \lambda_{\rm J} - 6 \varpi$ (top panel), $\sigma_{\rm S} = 20 \lambda - 
                  7 \lambda_{\rm S} - 13 \varpi$ (middle panel), and $\sigma_{\rm N} = \lambda - 2 \lambda_{\rm N} + \varpi$ (bottom panel) 
                  plotted against time for the time interval ($-$10, 30) kyr. The relative mean longitude with respect to Uranus appears as 
                  a thick blue line. The angle $\sigma_{\rm S}$ alternates between circulation and asymmetric libration, indicating that the 
                  motion is chaotic. The observed resonant evolution is consistent across control orbits.
                 }
         \label{mmrs}
      \end{figure}

     Transitions in and out or between the various co-orbital states are not triggered by encounters but result from complex multibody 
     ephemeral mean motion resonances as described in de la Fuente Marcos \& de la Fuente Marcos (2014). As other Uranian co-orbitals do, 
     2015~DB$_{216}$ moves in near resonance with the other three giant planets: 1:7 with Jupiter, 7:20 with Saturn, and 2:1 with Neptune. 
     Figure \ref{mmrs} shows the behaviour of the resonant arguments $\sigma_{\rm J} = 7 \lambda - \lambda_{\rm J} - 6 \varpi$, $\sigma_{\rm 
     S} = 20 \lambda - 7 \lambda_{\rm S} - 13 \varpi$, and $\sigma_{\rm N} = \lambda - 2 \lambda_{\rm N} + \varpi$, where $\lambda_{\rm J}$ 
     is the mean longitude of Jupiter, $\lambda_{\rm S}$ is the mean longitude of Saturn, $\lambda_{\rm N}$ is the mean longitude of 
     Neptune, and $\varpi = \Omega + \omega$ is the longitude of the perihelion of 2015~DB$_{216}$. The plot (similar to figs. 6 and 7 in de 
     la Fuente Marcos \& de la Fuente Marcos 2014) clearly indicates that transitions are quickly triggered when multiple mean motion 
     resonances work in unison. In Fig. \ref{mmrs}, an originally passing orbit becomes a horseshoe path after $\sigma_{\rm J}$ and 
     $\sigma_{\rm N}$ stop circulating; prior to the ejection from the horseshoe-like path, the same scenario is observed. The dynamical 
     role of three-body mean motion resonances has been recently explored by Gallardo (2014). Marzari et al. (2003) already pointed out that
     three-body resonances could be a source of instability for Uranian co-orbitals, in particular Trojans. 
%
%
%
%
     \hfil\par
     It may be argued that it is unclear from Fig. \ref{mmrs} that overlapping mean motion resonances are responsible for the transitions
     between co-orbital states or the activation/deactivation of the observed librational dynamics. On strictly theoretical grounds, this 
     is to be expected as the orbital architecture of the giant planets is not random. Ito \& Tanikawa (2002) and Tanikawa \& Ito (2007) 
     have pointed out that Jupiter affects the motions of Uranus and Neptune without the connection of Saturn and that secular perturbations 
     may be nullified in such context. To explore this issue further, we have recomputed the short-term orbital evolution of the nominal 
     orbit of 2015~DB$_{216}$ using increasingly complex physical models. Integrating the three-body problem ---Sun, Uranus, and 
     2015~DB$_{216}$--- we observe asymmetric horseshoe evolution with no transitions. Computing the evolution of the four-body problem 
     ---Sun, Jupiter, Uranus, and 2015~DB$_{216}$--- a transition from asymmetric horseshoe to L$_4$ Trojan at about 15 kyr is recorded. The 
     alternative four-body problem ---Sun, Saturn, Uranus, and 2015~DB$_{216}$--- results in an L$_4$ Trojan path with no transitions. A 
     similar result is observed for the case Sun, Uranus, Neptune, and 2015~DB$_{216}$. The six-body problem ---Sun, Jupiter, Saturn, 
     Uranus, Neptune, and 2015~DB$_{216}$--- results in an asymmetric horseshoe transitioning to a passing orbit, but somewhat earlier than 
     observed in Fig. \ref{mmrs}. It appears obvious that in order to turn the asymmetric horseshoe libration into a passing orbit, 
     superposition of mean motion resonances is required. 
     \hfil\par
     A more systematic exploration of the various five-, six- and higher-multiplicity-body problems shows that the details of the 
     transitions are strongly dependent on the number of distant perturbers included in the simulations. The dynamical evolution of these 
     objects is unusually sensitive to the physical model used to perform the calculations. Removing the three asteroids and the 
     Pluto-Charon system does not have a major observable impact on the outcome of the simulations both in terms of the timing and the 
     types of the observed transitions (the evolution displayed in Fig. \ref{mmrs} remains very nearly the same). However, stripping planets 
     from the model ---even Mercury or Mars--- has immediate effects on the orbital evolution of these recurring Uranian co-orbitals. For 
     example, removing Mercury from the calculations triggers a transition from asymmetric horseshoe to L$_4$ Trojan at about 15 kyr and 
     back to asymmetric horseshoe a few kyr later. This analysis indicates that any numerical study of these objects that is not using the
     full set of planets may arrive to unrealistic conclusions regarding the stability and dynamical evolution of these objects. This is
     consistent with the analysis in Tanikawa \& Ito (2007). Published works like those of Marzari et al. (2003) or Alexandersen et al. 
     (2013) made use of a five-body model including the Sun and the four outer planets.
     \hfil\par
\begin{figure}
       \centering
        \includegraphics[width=\linewidth]{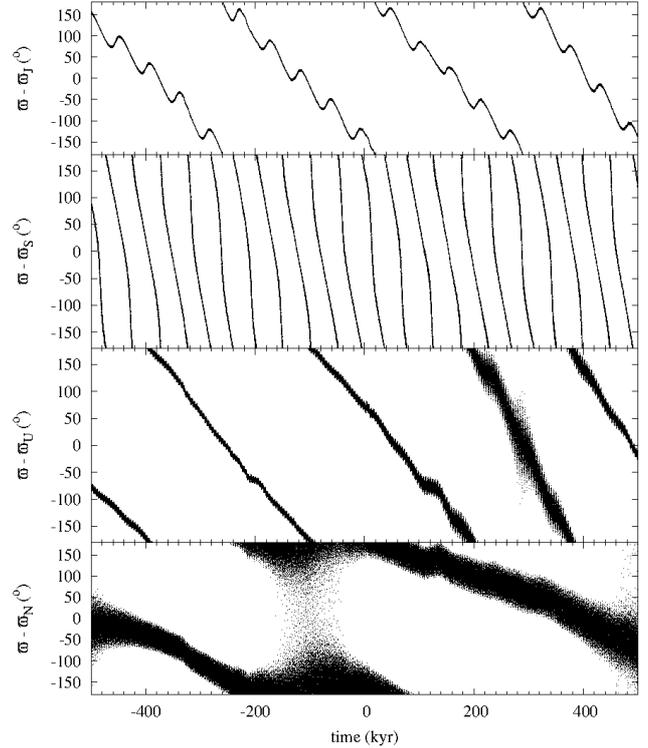}
        \caption{Time evolution of the relative longitude of the perihelion, $\Delta \varpi$, of 2015~DB$_{216}$ with respect to the giant 
                 planets: referred to Jupiter ($\varpi - \varpi_{\rm J}$), to Saturn ($\varpi - \varpi_{\rm S}$), to Uranus ($\varpi - 
                 \varpi_{\rm U}$), and to Neptune ($\varpi - \varpi_{\rm N}$). The relative longitudes circulate over the entire simulated 
                 period with the exception of that of Neptune. These results are for the nominal orbit in Table \ref{elements}.
                }
        \label{secular}
     \end{figure}

     As for the secular behaviour (see Fig. \ref{secular}), it is markedly different from the one described for other Uranian co-orbitals in
     de la Fuente Marcos \& de la Fuente Marcos (2014). The precession frequency of the longitude of the perihelion of 2015~DB$_{216}$, 
     $\varpi = \Omega + \omega$, is only in secular resonance with Neptune and for a limited time. The value of $\Delta \varpi = \varpi - 
     \varpi_{\rm N}$ librates around 180\degr. The absence of apsidal corotation resonances (see Lee \& Peale 2002; Beaug\'e, Ferraz-Mello 
     \& Michtchenko 2003) probably translates into increased stability of this object when compared with other Uranian co-orbitals. Even if 
     evidently chaotic, its dynamical evolution appears to be relatively stable and the object may remain in the neighbourhood of Uranus 
     co-orbital region for millions of years. On the other hand, during the quasi-satellite episode observed in Fig. \ref{controldb216}, 
     right-hand panels, the object exhibits Kozai-like dynamics with $\omega$ librating around 270\degr for about 100 kyr.     
%
%
%
%

  \section{Discussion}
     Our analysis suggests that, even if submitted to chaotic dynamics, this object may be intrinsically more stable than any of the 
     previously known Uranian co-orbitals, but there is an additional piece of robust evidence in favour of this interpretation. Asteroid 
     2015~DB$_{216}$ was serendipitously discovered by a survey aimed at finding near-Earth objects (NEOs), the Mt. Lemmon Survey (MLS), 
     that is part of the Catalina Sky Survey (CSS)\footnote{http://www.lpl.arizona.edu/css/css\_facilities.html} and precovered from 
     observations acquired by the SDSS,\footnote{http://www.sdss.org} a project aimed at creating the most detailed three-dimensional maps 
     of the Universe ever made after imaging about one-third of the sky. Therefore, its observation (past and present) was not the result of 
     careful planning like it was the case of the discovery of 2011~QF$_{99}$ (Alexandersen et al. 2013). In de la Fuente Marcos \& de la 
     Fuente Marcos (2014), section 8, we studied the discovery circumstances of known Uranian co-orbitals and candidates. All of them have 
     been found at declinations in the range $-2$\degr to +15\degr (see Table \ref{discovery}). In sharp contrast, 2015~DB$_{216}$ was 
     observed at declination +57\fdg3 by SDSS in 2003 and at +29\fdg5 by MLS in 2015. In de la Fuente Marcos \& de la Fuente Marcos (2014), 
     fig. 18, an observational bias regarding the observation of Uranian co-orbitals was pointed out, that co-orbitals reaching perigee (or 
     perihelion) near declination 0\degr are nearly six times more likely to be found than those reaching perigee at declinations 
     $\pm$60\degr, if they do exist. 
     \hfil\par
     Table \ref{discovery} includes data from the Minor Planet Center (MPC) Database\footnote{http://www.minorplanetcenter.net/db\_search} 
     and clearly shows that, among Uranian co-orbitals, 2015~DB$_{216}$ is a puzzling outlier. Assuming that this object is not a 
     statistical accident, its presence hints at the existence of a significant population of objects moving in similar orbits, perhaps an 
     order of magnitude larger than current models predict for regular Uranian co-orbitals. The absence of secular perturbations by Jupiter 
     and Uranus found for 2015~DB$_{216}$ may probably explain the relative stability of this putative population. It could be the case that 
     ---after all--- Uranus may host a large population of (transient but recurring) co-orbitals, but their orbits may be characterized by 
     high orbital inclinations. The discovery of 2015~DB$_{216}$ parallels that of 83982 Crantor (2002~GO$_{9}$), found by the Near-Earth 
     Asteroid Tracking (NEAT) project at Palomar Observatory in 2002 and precovered from images obtained in 2001 by the Air Force Maui 
     Optical and Supercomputing (AMOS) observatory and SDSS. 
%
%
      \begin{table}
        \centering
        \fontsize{8}{11pt}\selectfont
        \tabcolsep 0.15truecm
        \caption{Equatorial coordinates and apparent magnitudes (with filter) at discovery time of known Uranian co-orbitals and candidates
                 (J2000.0 ecliptic and equinox). Source: MPC Database.
                }
        \begin{tabular}{lccc}
          \hline
             Object                         & $\alpha$ ($^{\rm h}$:$^{\rm m}$:$^{\rm s}$) & $\delta$ (\degr:\arcmin:\arcsec) & $m$ (mag) \\
          \hline
             1999~HD$_{12}$                 & 12:31:54.80                                 & -01:03:07.9                      & 22.9 (R)  \\
             (83982) Crantor                & 14:10:43.80                                 & +01:24:45.5                      & 19.2 (R)  \\
             2002~VG$_{131}$                & 00:54:57.98                                 & +12:07:52.4                      & 22.5 (R)  \\
             2010~EU$_{65}$                 & 12:15:58.608                                & -02:07:16.66                     & 21.2 (R)  \\
             2011~QF$_{99}$                 & 01:57:34.729                                & +14:35:44.64                     & 22.8 (r)  \\
          \hline
             2015~DB$_{216}$ (SDSS, 2003)   & 08:29:42.21                                 & +57:19:08.2                      & 22.4 (V)  \\
             2015~DB$_{216}$ (MLS, 2015)    & 11:09:56.70                                 & +29:31:01.6                      & 20.5 (V)  \\
          \hline
        \end{tabular}
        \label{discovery}
      \end{table}
%
%
  \section{Exploring the orbital domain near 2015~DB$_{216}$}
     It could be debated that arguing on the existence of a population of high orbital inclination Uranian co-orbitals based solely on the 
     discovery of 2015~DB$_{216}$ is an exercise of mere speculation. In order to investigate this interesting hypothesis further, we have 
     studied the evolution of a sample of 10$^{3}$ fictitious bodies with initial orbits similar to that of 2015~DB$_{216}$. Their orbital 
     elements have been generated using uniformly distributed random numbers in order to survey the relevant region of the orbital parameter 
     space evenly. For each test orbit, a numerical integration for 10$^{4}$ yr ---using the same physical model and techniques applied in 
     previous sections--- has been performed. 
     \hfil\par
\begin{figure}
       \centering
        \includegraphics[width=\linewidth]{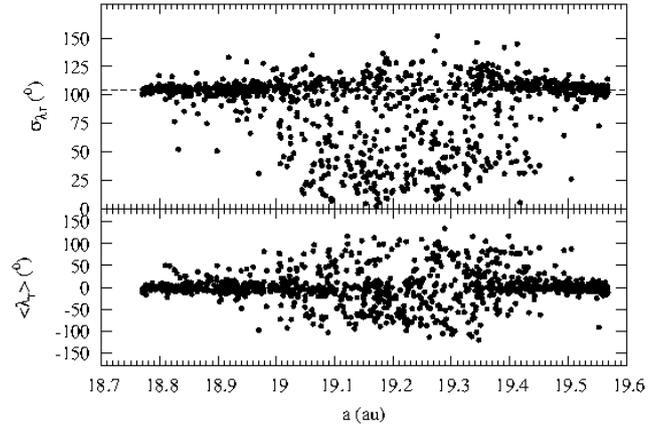}
        \caption{Mean value (bottom panel) and standard deviation (top panel) of the resonant angle, $\lambda_{\rm r}$, as a function of the 
                 initial value of the semimajor axis. The value of the standard deviation for a continuous uniform distribution of maximum 
                 value 180\degr and minimum value $-$180\degr is also indicated, $\sim$104\degr. 
                }
        \label{ala}
     \end{figure}
\begin{figure}
       \centering
        \includegraphics[width=\linewidth]{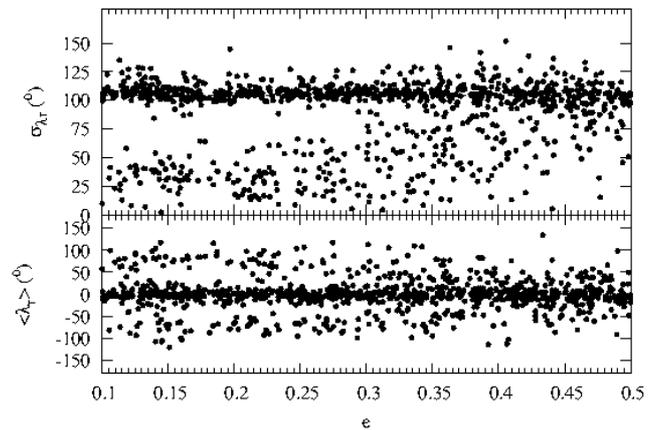}
        \caption{Same as Fig. \ref{ala} but for the initial value of the eccentricity.
                }
        \label{ela}
     \end{figure}
\begin{figure}
       \centering
        \includegraphics[width=\linewidth]{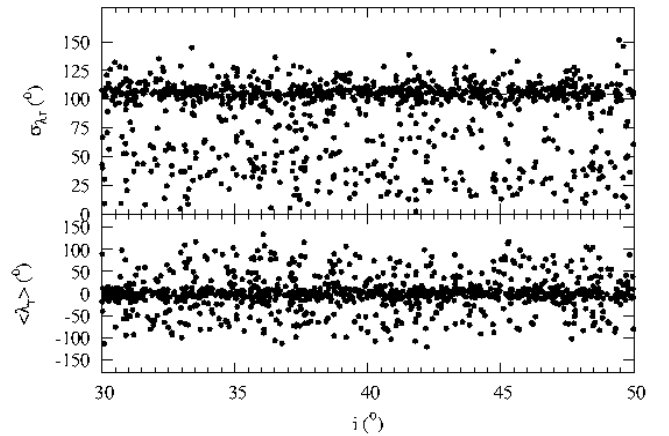}
        \caption{Same as Fig. \ref{ala} but for the initial value of the inclination.
                }
        \label{ila}
     \end{figure}

     The average value (bottom panels) of the resonant angle, $\lambda_{\rm r}$, and its standard deviation (top panels) as a function of 
     the initial values of the orbital parameters $a$, $e$ and $i$ is plotted in Figs \ref{ala}--\ref{ila}, respectively. The assumed ranges 
     in $a$, $e$ and $i$ are displayed; $\Omega$ and $\omega$ are chosen in the range 0\degr--360\degr. An object following a strict passing 
     orbit has an average value of the resonant angle close to 0\degr and its associated standard deviation is nearly 104\degr (also 
     displayed on the top panel of the figures as a dashed line); the value of the variance of a continuous uniform distribution of maximum 
     value $x_{\rm max}$ and minimum value $x_{\rm min}$ is given by the expression $(x_{\rm max}-x_{\rm min})^{2}/12$, and the mean value 
     is $(x_{\rm max}+x_{\rm min})/2$. Consistently with our analysis of the dynamics of 2015~DB$_{216}$ in which recurring co-orbital 
     episodes of various types are observed, the fictitious orbits studied here show values of the standard deviation of the resonant angle 
     in obvious conflict with those expected in a non-librational scenario. Therefore, there is a robust theoretical ground to assume that 
     such population of high orbital inclination Uranian recurring co-orbitals may exist.
%
%
%
%
%
%
%
%
%
%
%
%
     \hfil\par
     Figure \ref{ila} shows that the initial value of the orbital inclination does not have a major impact on $\langle\lambda_{\rm r}\rangle$ 
     or $\sigma_{\lambda_{\rm r}}$ (for the ranges of the values of the orbital elements considered here), but Fig. \ref{ela} indicates that 
     the initial value of the orbital eccentricity has a major influence on the subsequent evolution of the test orbit. For values in the 
     range 0.1--0.3 the magnitude of the standard deviation of the resonant angle tends to be significantly lower when recurring co-orbital 
     behaviour appears. These orbits are inherently more stable as their perihelia and aphelia are less directly perturbed. They are 
     associated with Trojan and quasi-satellite co-orbital states. 
     \hfil\par
\begin{figure}
       \centering
        \includegraphics[width=\linewidth]{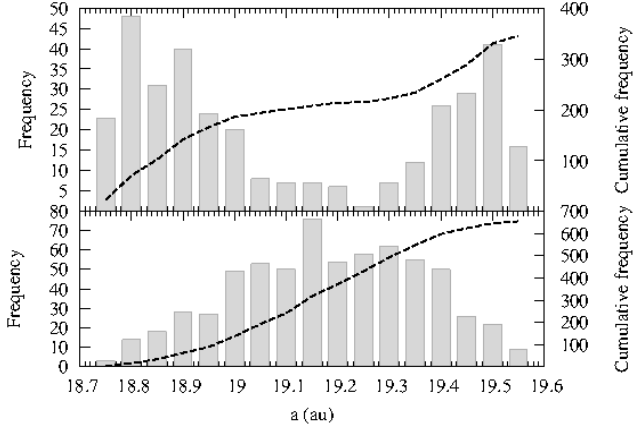}
        \caption{Distribution in initial semimajor axis for objects with standard deviation of the resonant angle, 
                 $\sigma_{\lambda_{\rm r}}$, in the range 100\degr--108\degr (likely passing orbits, top panel) and outside that range 
                 (bottom panel). Out of 10$^{3}$ orbits studied, 346 have $\sigma_{\lambda_{\rm r}}\in(100\degr, 108\degr)$.
                }
        \label{alast}
     \end{figure}
\begin{figure}
       \centering
        \includegraphics[width=\linewidth]{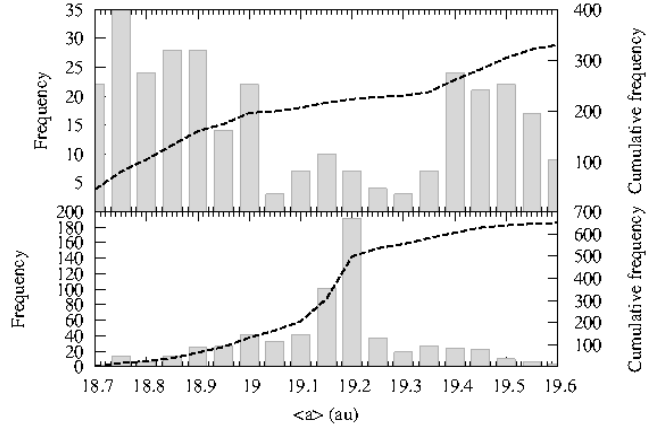}
        \caption{Same as Fig. \ref{alast} but for the average value of the semimajor axis. 
                }
        \label{alast2}
     \end{figure}

     Returning to the issue of the actual extension of the Uranian co-orbital zone for these high-inclination, transient but recurring 
     co-orbitals, the distribution in semimajor axis (the initial value) for test orbits with $\sigma_{\lambda_{\rm r}}\in(100\degr, 108)\degr$ 
     (top panel) and outside that range (bottom panel) is plotted in Fig. \ref{alast}. The distribution clearly shows that the co-orbital 
     region approximately goes from 19.0 to 19.4 au. Outside that range in semimajor axis, most trajectories become passing orbits. However, 
     even deep inside Uranus' co-orbital zone not all the values of the semimajor axis are equally favourable regarding stability. Figure 
     \ref{alast2} shows the distribution for the average value of the semimajor axis. The extension of the co-orbital zone is confirmed, but 
     those orbits with values of the osculating semimajor axis in the range 19.1--19.2 au are far more stable.
%
%
%
%
%
%
%
%
     \hfil\par
\begin{figure}
       \centering
        \includegraphics[width=\linewidth]{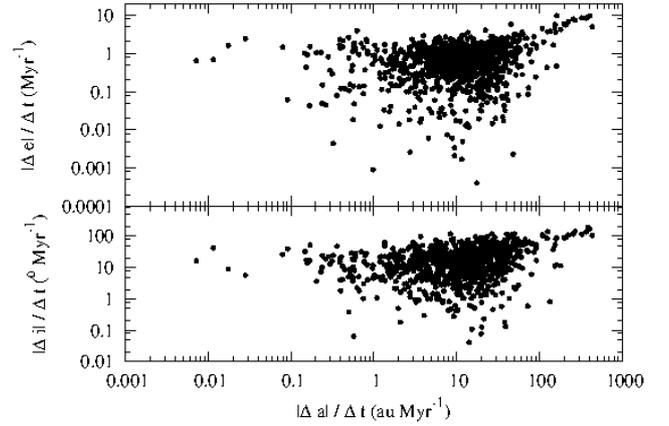}
        \caption{Variation of the eccentricity (top panel) and the inclination (bottom panel) over time as a function of the corresponding
                 variation in semimajor axis (see the text for details).
                }
        \label{drift}
     \end{figure}
\begin{figure}
       \centering
        \includegraphics[width=\linewidth]{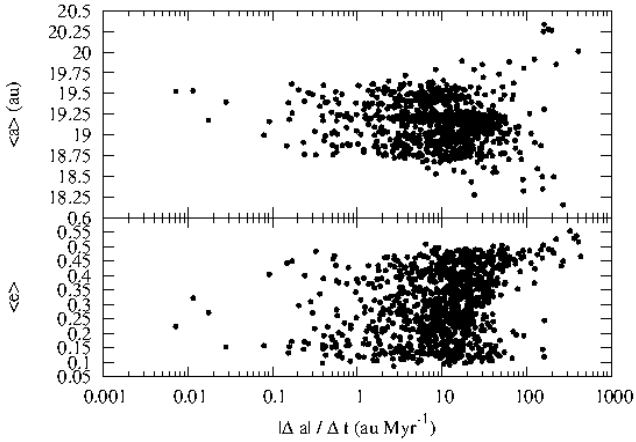}
        \caption{Average value of the semimajor axis (top panel) and the eccentricity (bottom panel) as a function of the variation of the 
                 semimajor axis over time.
                }
        \label{drift2}
     \end{figure}

     Although the previous analysis strongly suggests that 2015~DB$_{216}$ is not a statistical accident and that more temporary Uranian
     co-orbitals must exist at high orbital inclinations, our relatively short and non-extensive integrations appear to leave the question
     of long-term stability open. Can we expect that objects moving in high-inclination orbits like that of 2015~DB$_{216}$ will spend 10 or 
     100 Myr trapped in the 1:1 mean motion resonance with Uranus? Figs \ref{drift} and \ref{drift2} suggest an answer in the affirmative. 
     In order to study the variation over time of a given orbital parameter, we have computed the absolute value of the difference between 
     the initial and final values of the parameter and divided by the integrated time. Figure \ref{drift} shows these drifts in $a$, $e$ and
     $i$ per Myr. It is obvious that, taking into account the span of the co-orbital zone, relatively long-term stability is possible. 
     \hfil\par
     Figure \ref{drift2} shows the average values of $a$ and $e$ as a function of the drift in $a$. From there, the most stable test orbits 
     are found for the ranges in $\langle{a}\rangle$ and $\langle{e}\rangle$ of 19.0--19.6 au and 0.15--0.35, respectively. The range in 
     $\langle{a}\rangle$ appears to be somewhat in conflict with the values found above, but there is an additional type of co-orbital 
     motion that does not require libration of the resonant angle: minor bodies following passing orbits with small Jacobi constants but 
     still moving in unison with a host planet as described by Namouni (1999). This orbital regime is also known as the Kozai domain because 
     it corresponds to a Kozai resonance (Kozai 1962). Under the Kozai resonance, both eccentricity and inclination oscillate with the same 
     frequency but out of phase; when the value of the eccentricity reaches its maximum the value of the inclination is the lowest and vice 
     versa ($\sqrt{1 - e^2} \cos i \sim$ constant); therefore, relatively large oscillations in $e$ and $i$ are still compatible with 
     long-term stability in this case. The most stable test orbit generated in our exploratory calculations could remain virtually unchanged 
     for time-scales well in excess of 10 Myr (see Fig. \ref{drift2}) and it is not a classical (librating) co-orbital but a fictitious 
     object in the Kozai domain. The actual values of the semimajor axis, 19.2 au, and eccentricity, 0.32, of 2015~DB$_{216}$ place this 
     object in the most stable region of the orbital domain probed in Fig. \ref{drift2}. If such orbits represent nearly 0.3 per cent of the 
     ones explored in this section and one actual object has already been found, 2015~DB$_{216}$, a significant number of less stable 
     high-inclination, recurring Uranian co-orbitals are likely to exist.
%
%
%
%
%
%
%
%

  \section{Conclusions}
     In this paper, we have analysed the orbital behaviour of 2015~DB$_{216}$ that is the fourth known minor body to be trapped in a 
     1:1 mean motion resonance with Uranus. Our numerical integrations show that it currently moves in a complex, horseshoe-like orbit when 
     viewed in a frame of reference corotating with Uranus. The object exhibits resonance angle nodding as it undergoes recurrent 
     co-orbital episodes with Uranus. Its high orbital inclination clearly separates this object from the other three known Uranian 
     co-orbitals and makes it more stable. Its discovery circumstances also single this minor body out among objects currently trapped into 
     the 1:1 commensurability with Uranus, hinting at the presence of a large number of similar objects. If they are inherently more stable 
     at higher inclinations, that should have an impact on the population of Uranian irregular moons. All but one are retrograde and their 
     orbital inclinations are in the range 139\degr--167\degr or in prograde terms 41\degr--13\degr. Five irregular moons out of 9 have 
     inclinations in the range 139\degr--147\degr. As for the mechanism responsible for the activation of the co-orbital states, 
     multibody mean motion resonances trigger the transitions as previously observed for other Uranian co-orbitals. 

  \section*{Acknowledgements}
     We thank the anonymous referee for his/her constructive and helpful report, and S.~J. Aarseth for providing the code used in this 
     research. In preparation of this paper, we made use of the NASA Astrophysics Data System, the ASTRO-PH e-print server, and the MPC data 
     server.

  \bsp
  \label{lastpage}
\end{document}